# A Model Obfuscation and Controllable Authorization Scheme for Deep Learning as a Service

**Nianyun Song[1,2,3], Haodi Wang[4], Yu Guo[1,2,3], Xiaohua Jia[4]**

[1] School of Artificial Intelligence, Beijing Normal University, Beijing 100875, China
[2] Engineering Research Center of Intelligent Technology and Educational Application, Ministry of Education, Beijing 100875, China
[3] Beijing Key Laboratory of Artificial Intelligence for Education, Beijing 100875, China
[4] Department of Computer Science, City University of Hong Kong, Hong Kong 999077, China

**Abstract.** Developing deep learning models requires substantial computational and data resources, making them core intellectual assets for enterprises. As Deep Learning as a Service (DLaaS) increasingly extends to edge computing environments, models deployed on third-party devices face growing threats from reverse engineering. Existing model protection methods often depend on specialized hardware or incur substantial computational overhead, while lacking effective mechanisms for controlling model usage. There is therefore a need for lightweight solutions that combine model protection with controllable authorization. To address this challenge, we propose KeyObf, a model obfuscation and controllable authorization scheme for DLaaS. KeyObf efficiently protects model parameters and architecture through model obfuscation while binding parameter recovery to user keys, ensuring that only authorized users can perform inference correctly. By integrating a key-based mechanism into the obfuscation framework, KeyObf unifies model protection and controllable authorization. Experimental results demonstrate that KeyObf preserves the inference accuracy of authorized users, rejects inference requests when credential authentication fails, and increases the difficulty of recovering model information under the evaluated reverse-engineering conditions. These results demonstrate the feasibility of a lightweight approach to model protection and controllable authorization in third-party deployment scenarios.

## 0 Introduction

Deep learning has achieved remarkable success in fields such as image recognition, natural language processing, and autonomous driving, driving the widespread adoption of deep learning models [1]. Initially, models were primarily deployed in the cloud, where users accessed inference services through application programming interfaces (APIs). However, to meet requirements for low latency, offline availability, and data privacy, an increasing number of models have been deployed directly on mobile devices and Internet of Things (IoT) terminals or distributed to third-party systems as software components, giving rise to a new application paradigm of Deep Learning as a Service (DLaaS). In this process, models have gradually evolved from cloud-hosted algorithms into deliverable and reusable digital assets, embedded in various applications and systems to perform local inference tasks. This evolution has substantially expanded their application scope and enhanced their commercial value.

Despite these benefits, this deployment paradigm also significantly increases model security risks. Developing high-performance deep learning models typically requires extensive private datasets, substantial computational resources, and prolonged optimization by domain experts, making trained models valuable core assets for enterprises. However, once these high-value models are distributed to third-party devices in the form of Android application packages (APKs), firmware, or container images, attackers can extract model files at relatively low cost using reverse-engineering tools, thereby obtaining model parameters and architectural information [2][3]. In split inference scenarios, Fan et al. [4] demonstrated that intermediate activations may still leak information about user inputs. To address these risks, researchers have proposed various protection methods based on different technical approaches. Trusted Execution Environment (TEE)-based solutions employ hardware isolation mechanisms, such as Intel SGX, to execute model inference within secure enclaves, reducing information leakage during model loading and execution [5]. Cryptographic approaches, including homomorphic encryption and secure multi-party computation, protect the privacy of model parameters and user inputs without requiring trusted hardware. Among these approaches, Gilad-Bachrach et al. [6] introduced CryptoNets, pioneering neural network inference based on homomorphic encryption. Subsequent studies [7][8] further explored efficient privacy-preserving machine learning using secure multi-party computation. As a passive defense and ownership verification mechanism, model watermarking embeds verifiable identifiers into model parameters or output behavior to provide evidence of model ownership after infringement occurs. Uchida et al. [9] were the first to introduce a parameter regularization term during training to embed watermarks into convolutional layer weights, laying the foundation for white-box model watermarking.

Although these approaches provide protection from different perspectives, each has limitations in third-party deployment scenarios. TEE-based solutions rely on specific hardware support and are constrained by limited secure memory capacity, reducing deployment flexibility across heterogeneous edge devices. Cryptographic approaches can protect the confidentiality of model parameters, but the computational graph structure remains exposed, and their computational overhead can be several orders of magnitude higher than that of plaintext inference. Model obfuscation protects model information by transforming parameters, operators, and network structures without relying on dedicated trusted hardware, offering a lightweight approach to protecting models deployed on edge devices [10][11]. However, existing obfuscation-based protection schemes have paid relatively little attention to controlling the authorization of inference capabilities. In third-party deployment scenarios, once a model has been delivered, recipients can directly invoke its inference interface and obtain valid outputs, regardless of whether they possess legitimate usage rights. Consequently, models may be used beyond the agreed scope or redistributed to unauthorized parties. Although model watermarking can provide evidence of ownership after infringement, it remains a passive tracing mechanism that cannot prevent unauthorized use in advance. Existing active authorization methods either require model retraining or rely on trusted hardware [12].

To address these limitations, this paper proposes KeyObf, a model obfuscation and controllable authorization scheme for DLaaS. Starting from model obfuscation, KeyObf extends existing parameter and architecture obfuscation techniques by further linking correct model inference to user authorization. Specifically, the proposed scheme applies scaling transformations to model parameters through coupling transformations and splits the recovery coefficient at each coupling position into an embedded factor and a key factor, making correct computational compensation dependent on the key factor obtained through authorization. It further combines operator and architecture obfuscation to increase the difficulty of model reverse engineering and introduces an independent user authorization credential mechanism that supports the addition of authorized users without modifying the deployed model. Finally, we implement the proposed scheme on TensorFlow Lite (TFLite), a widely used edge inference framework, and evaluate it in terms of model functional consistency, authorization effectiveness, resistance to reverse-engineering attacks, and runtime overhead. We also compare its performance with existing model obfuscation methods.

# 1 Related Work

To address model protection challenges, researchers have proposed various approaches based on different technical foundations. TEE-based approaches protect sensitive computations through hardware isolation, with Intel SGX and ARM TrustZone being two widely used implementations. Ohrimenko et al. [5] pioneered privacy-preserving computation over data from multiple parties within an SGX environment. Islam et al. [13] proposed a confidential deep learning inference system based on ARM TrustZone, which loads and decrypts model parameters within the TEE to provide confidentiality and integrity protection for models and data on resource-constrained edge devices. However, this approach is constrained by the limited secure memory available in the TEE. Zhang et al. [14] investigated the security of TEE-based deep neural network (DNN) partitioning schemes and revealed their vulnerability to attackers with knowledge of pretrained models. Li et al. [15] further proposed TEESlice, which employs a partition-before-training strategy to effectively resist such attacks. Nevertheless, TEE-based approaches rely heavily on specific hardware, limiting deployment flexibility across heterogeneous devices.

Cryptographic protection methods primarily follow two approaches: homomorphic encryption and secure multi-party computation. Homomorphic encryption enables computations to be performed directly on ciphertexts, allowing model inference without decryption. Gilad-Bachrach et al. [6] proposed CryptoNets, the first application of homomorphic encryption to neural network inference, which employed the YASHE scheme to classify encrypted images. Lee et al. [16] were the first to introduce bootstrapping into privacy-preserving machine learning, enabling encrypted inference with ResNet-20 using the RNS-CKKS scheme. However, the computational overhead of homomorphic encryption remains several orders of magnitude higher than that of plaintext inference. Secure multi-party computation enables multiple parties to jointly perform computations without revealing their private inputs. Mohassel et al. [7] proposed the SecureML framework, while Wagh et al. [17] introduced the SecureNN protocol to improve the efficiency of neural network inference. Related techniques have since been extended to robustness verification [18], medical image segmentation [19], privacy-preserving neural architecture search [20][21], and architecture-hiding inference [22]. Nevertheless, secure multi-party computation requires the online participation of multiple parties and incurs substantial communication overhead, making it unsuitable for scenarios in which models are directly distributed to third-party devices.

Model watermarking embeds verifiable identifiers into models to provide evidence of ownership after infringement occurs. Uchida et al. [9] introduced digital watermarking into deep neural networks by incorporating a parameter regularization term to embed watermarks during training. Zhang et al. [23] and Adi et al. [24] proposed backdoor-trigger-based black-box watermarking schemes that enable ownership verification without accessing model parameters. However, watermarking is inherently a passive defense mechanism. It can provide supporting evidence after infringement but cannot prevent unauthorized model reuse or misuse. Moreover, watermarking remains vulnerable to watermark detection, removal, and forgery attacks [25].

Model obfuscation increases the difficulty of reverse engineering by introducing complexity into model structures and representations while preserving functional equivalence. Zhou et al. [10] proposed ModelObfuscator, a tool that protects on-device models through operator renaming, parameter packaging, neural structure obfuscation, and shortcut connection injection. Xu et al. [26] proposed DeepObfuscation, which employs knowledge distillation to compress a complex feature extractor into a shallow network, thereby concealing information about the original architecture. Zhou et al. [11] further introduced a dynamic obfuscation strategy that uses coupled weight transformations to keep intermediate results obfuscated throughout inference, effectively resisting dynamic instrumentation attacks. However, these obfuscation methods primarily focus on protecting model parameters and architectures without providing an independent user authorization mechanism. Zhou et al. [12] proposed NNSplitter, which selectively perturbs model weights to degrade inference performance in the unauthorized state and uses a TEE to protect the secret information required for model recovery, thereby enabling active control over model usage. However, this approach relies on trusted hardware.

In summary, existing model protection methods remain limited in their ability to provide both lightweight model protection and usage authorization, restricting the secure deployment of deep learning models in third-party environments. To address these limitations, this paper builds upon existing model obfuscation techniques by binding the recovery coefficients of coupling transformations to user keys, thereby achieving both model protection and usage authorization without relying on trusted hardware.

# 2 Problem Statement

## 2.1 System Architecture

The system model consists of two main entities: the model owner and authorized users. Their responsibilities are described as follows.

(1) **Model Owner.** The model owner is the developer and intellectual property holder of the deep learning model. The model owner invests substantial computational resources and domain expertise in designing the model architecture and training its parameters to meet the performance requirements of practical applications. For commercial deployment, the model owner needs to distribute the trained model to authorized third-party users. Before distribution, the model owner applies the proposed obfuscation scheme to protect the model against reverse engineering and unauthorized use. Specifically, the original model undergoes parameter, operator, and architecture obfuscation to produce a protected model. The model owner also generates a dedicated key and an independent authorization credential for each user.

(2) **Authorized User.** An authorized user is a user who has been granted permission to use the model. The user obtains the protected model, the corresponding key, and an authorization credential from the model owner and deploys them on a local device or server. During inference, the user submits the input data together with the key and authorization credential. The inference library then authenticates the credential and performs the compensation operations for the coupling layers. Only users possessing a valid key and its corresponding authorization credential can obtain correct inference outputs. Requests that fail authentication are rejected.

### 2.2 Threat Model

The threat model considered in this paper is defined as follows.

(1) **Model Owner.** The model owner is a fully trusted party responsible for performing model obfuscation, generating the master key, and distributing user keys and independent authorization credentials to authorized users through secure channels. The model owner does not intentionally disclose keys or information about the original model. This assumption is reasonable because, as the intellectual property holder, the model owner has the strongest incentive to protect the model.

(2) **Authorized User.** Authorized users are considered semi-honest parties. On the one hand, they follow the authorization protocol, correctly use their assigned keys to perform model inference, and do not intentionally share their keys or authorization credentials with others. On the other hand, motivated by curiosity or commercial competition, authorized users may passively analyze the received obfuscated model, including through model format conversion and buffer parsing, in an attempt to infer its architecture and parameters.

(3) **Unauthorized Adversary.** Unauthorized adversaries are considered malicious. An adversary may obtain an obfuscated model file through illicit means but cannot obtain a valid user key or authorization credential. Under these conditions, the adversary may attempt to steal the model's intellectual property or bypass the authorization mechanism through techniques such as model format conversion, reverse parsing of buffers, and feature-behavior analysis. Given that inference programs deployed on third-party devices may be subject to local reverse engineering and tampering, we do not assume that the authorization authentication module cannot be bypassed. To evaluate the protection provided by the key-binding mechanism when authentication is compromised, we further simulate black-box searches for candidate key factors in a controlled environment. The experimental settings are detailed in Section 4.2.

### 2.3 Design Goals

Based on the threat model described above, the proposed scheme aims to achieve the following design goals:

(1) **Model Protection.** The obfuscation scheme should effectively protect model assets, including model parameters and network architectures, by increasing the difficulty for attackers to recover original model information through reverse engineering.

(2) **Usage Authorization.** The scheme should provide control over model usage permissions. Users possessing valid keys and authorization credentials should be able to perform correct inference, while requests that fail authentication should be rejected.

(3) **Preservation of Inference Accuracy.** The obfuscation process should not compromise the model's inference capabilities. When authorized users provide the correct keys, the obfuscated model should produce inference results consistent with those of the original model for the same inputs, thereby preserving functional integrity.

(4) **Practicality and Efficiency.** The additional computational overhead introduced by the scheme should remain within acceptable limits, enabling efficient inference on third-party devices and meeting practical deployment requirements.

## 3 Scheme Design

### 3.1 Overview

The proposed scheme consists of three stages: model obfuscation, user authorization, and authorized inference. (1) **Model obfuscation** is performed by the model owner as a one-time offline operation, with the primary objective of transforming the original model into a protected obfuscated model. In this stage, parameter obfuscation is first applied to perturb the original weight distribution. The recovery coefficients are then split into embedded factors and key factors through a key-binding mechanism. Operator and architecture obfuscation are subsequently applied to conceal model information, followed by model packaging and compilation to produce a deployable obfuscated model. (2) **User authorization** is performed by the model owner as needed. A user-specific key $UK$ is derived from the master key $MK$, a random salt $Salt$, and the user identifier $UID$, and an independent authorization credential is issued. This process does not require any modification to the deployed model. (3) **Authorized inference** is performed by an authorized user. The user provides private input data $X$, the user identifier $UID$, the user key $UK$, and the authorization credential to the obfuscated model. The inference library internally performs credential authentication and decryption, and uses the recovered key factors to perform coupling compensation. If authentication fails, the inference request is rejected.

These mechanisms serve distinct but complementary roles in model protection and usage authorization. Coupling-based parameter obfuscation provides recovery coefficients that can be split for key binding, making correct model compensation dependent on key factors obtained through authorization. Operator renaming, dimension obfuscation, topology perturbation, and parameter packaging further increase the difficulty of reverse engineering by obscuring operator semantics, network structures, and parameter storage. The custom inference library integrates the execution of obfuscated operators with credential authentication and decryption, as well as coupling compensation, thereby enabling model information protection and usage authorization to work together.

## 3.2 Model Obfuscation

### 3.2.1 Parameter Obfuscation

For linear operators, parameter obfuscation can be achieved by applying reversible scaling transformations to weights and biases. Consider a fully connected layer, where $X_{i-1}$ and $X_i$ denote the input and output vectors of the $i$-th layer, respectively, and $W_i$ and $b_i$ denote its weight matrix and bias vector. The original computation is given by $X_i = W_i^T X_{i-1} + b_i$. With the input unchanged, multiplying both the weights and biases by a scaling factor $\alpha$ yields $X_i' = (\alpha W_i)^T X_{i-1} + \alpha b_i = \alpha X_i$, scaling the output by a factor of $\alpha$. By exploiting the propagation property of this linear transformation, the accumulated scaling factor can be cancelled through a $1/\alpha$ transformation at a subsequent coupling layer. Lemma 3.1 specifies the transformation rules for purely linear paths.

**Lemma 3.1 (Coupling Transformation for Linear Paths).** Let a subnetwork $f$ consist of $n$ consecutive linear layers ($n \geq 2$), where the $i$-th layer computes $X_i = W_i^T X_{i-1} + b_i$, for $i \in [1, n]$. For any $\alpha > 0$, the network parameters are transformed as follows:

$$\begin{gathered} W_1' = \alpha W_1, \\ W_n' = \frac{1}{\alpha} W_n, \\ b_i' = \alpha b_i \left(i \in [1, n-1]\right) \end{gathered} \tag{1}$$

All other parameters, including $b_n$, remain unchanged. The transformed network $f'$ then satisfies $f'(X_0) = f(X_0)$.

In this transformation, the layer to which the perturbation is applied is referred to as the *selected layer*, while the layer that cancels the perturbation is referred to as the *coupling layer*. These layers are paired to preserve the network output. Lemma 3.1 requires a purely linear path between the selected layer and the coupling layer. However, in practical models, linear layers are often followed by nonlinear activation functions such as ReLU. Moreover, frameworks such as TFLite frequently fuse linear and activation layers into a single operator, making it difficult to insert a coupling point between them. To accommodate nonlinear activation layers commonly found in deep learning models, such as $ReLU_\beta$, we extend the coupling transformation approach in [11] and develop a coupling mechanism involving clipped ReLU activation based on Theorem 3.2. This extension aims to generalize the obfuscation method from purely linear paths to nonlinear computational paths that satisfy specific transformation conditions. By exploiting the clipping property of $ReLU_{\beta}$, the mechanism inserts a coupling layer with a weight of $1/\alpha$ after the activation layer and applies clipped ReLU again, enabling exact restoration of the scaled feature values.

**Theorem 3.2 (Coupling Transformation for Nonlinear Paths).** Let $\beta \geq 0$ be the clipping threshold, and define $ReLU_{\beta}(x)$ as

$$ReLU_{\beta}(x) = \begin{cases} \beta, & x \geq \beta, \\ x, & 0 < x < \beta, \\ 0, & \text{otherwise}. \end{cases}$$

Let a subnetwork $f$ consist of $n$ consecutive linear layers, where the $i$-th layer computes $X_i = W_i^T X_{i-1} + b_i$, for $i \in [1, n]$, with no nonlinear activations or other operations between layers that would disrupt the scaling relationship. For a scaling factor $\alpha \in (0, 1)$, apply the following parameter transformations:

$$W_1{}' = \alpha W_1, W_i{}' = W_i (2 \leq i \leq n), b_i{}' = \alpha b_i (1 \leq i \leq n).$$

The resulting obfuscated subnetwork $f'$ satisfies $f'(X_0) = \alpha f(X_0)$. Furthermore, after inserting a linear coupling operator with weight $\frac{1}{\alpha} I$ at the output, we obtain

$$ReLU_{\beta}\left(\frac{1}{\alpha} I_d ReLU_{\beta}(f'(X_0))\right) = ReLU_{\beta}(f(X_0)) \tag{2}$$

Here, $I_d$ is the identity matrix whose dimension matches that of the output features.

Implementing the coupling transformation requires determining the pairing between selected layers and coupling layers. Specifically, when the path between a selected layer and a subsequent layer is purely linear, the subsequent layer can directly serve as the coupling layer according to Lemma 3.1. If the path terminates with a clipped ReLU activation, the extended mechanism in Theorem 3.2 can be applied by inserting an additional linear layer after the activation, provided that the scaling propagation conditions are satisfied. The inserted layer has a weight matrix of $\frac{1}{\alpha} I$ and a zero bias vector, followed by the same clipped $ReLU_{\beta}$ operation to satisfy the conditions of Theorem 3.2. For residual connections, the coupling layer can be placed within a residual branch to restore the features before branch merging. Alternatively, it can be placed on the main residual path, in which case all branches involved in the summation must use consistent scaling. For BatchNorm, the operation can be fused with an adjacent linear layer during inference, after which the transformation is applied to the equivalent parameters. For fused activation operators, coupling-based restoration is performed after the activation computation. No corresponding transformation is applied to computational paths that do not satisfy these conditions. In practice, multiple pairs of selected layers and coupling layers are randomly chosen and assigned different scaling factors, while ensuring that the combined transformations preserve overall functional equivalence. This increases the difficulty of reverse engineering. In addition, the newly inserted coupling layers perform actual parameter compensation while modifying the model's computational graph, further increasing the difficulty of reverse engineering when combined with topology perturbation.

### 3.2.2 Key Binding

To enable usage authorization control, we extend the coupling transformation described above as follows. Let $\alpha_j$ denote the scaling factor of the $j$-th layer pair. We split the recovery coefficient $1/\alpha_j$ at the $j$-th coupling position into two components: an embedded factor $r_{emb,j}$ and a key factor $r_{key,j}$, satisfying

$$\frac{1}{\alpha_j} = r_{emb,j} \cdot r_{key,j}. \tag{3}$$

The embedded factor is statically incorporated into the weights of the coupling layer during model obfuscation, whereas the key factor is derived by the model owner during the same stage and is not included in the deployment package. During inference, the key factor is obtained by authenticated decryption of the user's independent authorization credential. In the model obfuscation stage, the model owner first generates a master key $MK$ and a salt $Salt$:

$$MK \xleftarrow{R} (256), Salt \xleftarrow{R} (128).$$

Here, $R(n)$ denotes the generation of an $n$-bit cryptographically secure random value. The master key $MK$ serves as the root key of the authorization system and is used to derive both the key factors and user-specific keys. It is kept secret by the model owner. The salt $Salt$ serves as a model-level randomization parameter to distinguish key derivation results across different models. Key factors are derived from the master key using HMAC-SHA256, which serves as a pseudorandom function. The derivation is defined as

$$r_{key,j} = MapToFloat\left(HMAC_{MK}\left(Encode\left(Salt, \text{“keyfactor”}, j\right)\right)\right). \tag{4}$$

Here, $j$ identifies the coupling position; $Encode(\cdot)$ denotes an unambiguous, length-prefixed, fixed-length encoding; and $HMAC_{MK}(\cdot)$ denotes HMAC-SHA256 keyed with the master key $MK$. The domain-separation label “keyfactor” ensures cryptographic separation between these factors and keys derived for other purposes. The function $MapToFloat(\cdot)$ takes the first 64 bits of the hash output and shifts them right by 11 bits to obtain a 53-bit unsigned integer $u$. It then applies the linear mapping $r = a + (b-a) \cdot \frac{u}{2^{53}}$ to map the integer to the interval $[a, b)$ for subsequent compensation computations. According to Eq. (3), the embedded factor $r_{emb,j}$ is determined by the recovery coefficient and the key factor:

$$r_{emb,j} = \frac{1}{\alpha_j \cdot r_{key,j}}. \tag{5}$$

The embedded factor replaces the complete recovery coefficient of the original coupling layer and is incorporated into its weights. Consequently, the obfuscated model contains only the embedded factor, rather than the complete recovery coefficient $1/\alpha_j$. If the obfuscated model is executed directly without using the authorization interface, the coupling layer cannot complete the compensation because the key factor is unavailable.

The key-binding mechanism described above operates at the granularity of coupling positions but can be further refined to individual output features, corresponding to output channels in convolutional layers and output neurons in fully connected layers. Suppose that the $j$-th coupling position contains $c_j$ output features. An independent key factor can be assigned to each output feature:

$$r_{key,j,k} = MapToFloat\left(HMAC_{MK}\left(Encode\left(Salt, \text{“keyfactor”}, j, k\right)\right)\right), k = 1, \ldots, c_j.$$

The corresponding embedded factors are configured to satisfy

$$r_{emb,j,k} \cdot r_{key,j,k} = \frac{1}{\alpha_j}.$$

All output features at the same coupling position share the scaling factor $\alpha_j$, and the two factors are paired to complete compensation before any cross-feature mixing occurs. Under per-feature configuration, the total number of key factors is

$$D = \sum_{j=1}^{m} c_j,$$

where $m$ denotes the number of coupling positions.

### 3.2.3 Operator and Architecture Obfuscation

Despite parameter obfuscation, attackers may still infer operator semantics from operator names or identify known model architectures by analyzing network topology. Following the approach in [10], this section introduces protection mechanisms at two levels: operator naming and network architecture.

**Operator Renaming.** This step conceals model structure information by replacing operator names. Let the original model contain an operator set $O=\{o_1, o_2, \ldots, o_n\}$. The renaming transformation maps this set to $O'=\{o'_1, o'_2, \ldots, o'_n\}$, where $o_i'$ is a random string of length $k$. Specifically, the system generates a unique random identifier for each operator instance. For example, the operators "Conv2D", "BatchNorm", and "ReLU" may be renamed to "wripyx", "mjzdmh", and "fsebwa", respectively. Notably, different instances of the same operator type are assigned distinct random names to further strengthen obfuscation. After replacing the operator names, the corresponding operator implementations in the inference engine must also be modified by replacing their original identifiers with the obfuscated ones. The modified inference engine is then recompiled to produce a custom inference engine capable of correctly recognizing and executing the renamed operators.

**Architecture Obfuscation.** Obfuscating operator names and parameters alone is insufficient to protect a model, as attackers may still infer its functionality or identify similar architectures by examining the network structure. For example, if the input, hidden, and output layers contain 4, 2, and 1 nodes, respectively, an attacker may use this information to search public model repositories for substitute models. To address this vulnerability, the proposed scheme protects model architectures through dimension obfuscation and topology perturbation.

Dimension obfuscation modifies the output dimensions of individual layers recorded in the model file. Let the original output shapes be represented by $s=(s_0, s_1, \ldots, s_n)$, where $s_i$ denotes the output dimension of the $i$-th layer. The proposed scheme provides two obfuscation strategies. The random strategy generates random shapes $r=(r_0, r_1, \ldots, r_n)$ to replace the original values. The alignment strategy determines the maximum shape $s_{max}=max(s_0, s_1, \ldots, s_n)$ and uniformly pads the dimensions of all layers to $s_{max}$. Since the modified TFLite library does not rely on the shape declarations stored in the model file during inference, this obfuscation does not affect model performance.

Topology perturbation disrupts the original network structure by injecting dummy components. Let the original computational graph be $G=(V, E)$. The proposed scheme introduces a set of dummy edges $E_{fake}$ and a set of dummy nodes $V_{fake}$, constructing an obfuscated graph $G'=(V \cup V_{fake}, E \cup E_{fake})$. Specifically, dummy edges are injected through skip connections: a pair of layers $(l_i, l_j)$ is randomly selected, and the output index of $l_i$ is added to the input list of $l_j$. Dummy nodes are introduced as redundant layers by inserting additional computational units between selected layer pairs. Each inserted unit receives its input from the preceding layer, and its output is added to the input list of the subsequent layer. Because the outputs of these dummy components are not used by the actual layers, they do not affect inference results. Unlike the coupling layers, which perform actual compensation computations, these redundant components do not participate in effective computation and are introduced solely to increase the difficulty of analyzing the model topology.

### 3.2.4 Model Packaging and Compilation

After the preceding transformations, the obfuscated model structure and parameters are further packaged and compiled to produce a deployable model.

**Parameter Packaging.** Standard on-device model files, such as TFLite files, store parameters in plaintext, allowing weights and biases to be directly parsed and extracted. The proposed scheme removes the obfuscated parameters $W'$ and $b'$ from the model file and embeds them as hard-coded constants in the operator implementation code of the inference library. For example, a layer originally computing $Y=\sigma(W^T X+b)$ is encapsulated as $Y=g(X)$, with its parameter values compiled into the binary along with the code. After this transformation, the model file retains only the topology description, and extracting the parameters requires binary reverse engineering of the inference library, substantially increasing the cost of analysis.

**Model Assembly and Library Compilation.** The obfuscated computational graph is assembled into a new model file. Because the operators have been renamed using random strings, the original inference library can no longer recognize them. Taking TFLite as an example, the source code implementations of the original operators are copied, their operator identifiers are replaced with the corresponding obfuscated identifiers, and the modified code is recompiled to generate a custom inference library. The compiled library can correctly recognize and execute the obfuscated operators. Finally, the obfuscated model file and the custom inference library together constitute a deployable protected model. Furthermore, code obfuscation can be applied to the custom inference library to enhance its resistance to reverse-engineering attacks.

## 3.3 User Authorization

The user authorization stage is performed by the model owner, who generates a user-specific key and issues an independent authorization credential for each legitimate user. This process involves only lightweight cryptographic operations and does not require any modification to the deployed model. Based on the HMAC-SHA256 mechanism described in Section 3.2.2, let $UID_i$ denote the identifier of the $i$-th user. The corresponding user key $UK_i$ is derived as follows:

$$UK_i = HMAC_{MK}\left(Encode\left(Salt, \text{“userkey”}, UID_i\right)\right). \tag{6}$$

Here, the string “userkey” serves as a domain-separation label that distinguishes different key purposes, ensuring that user-key derivation is independent of key-factor derivation in Eq. (4). Assuming that HMAC-SHA256 acts as a pseudorandom function, this construction provides three security properties: (1) **Master-key confidentiality:** Knowledge of multiple user keys $UK_i$ and their corresponding identifiers $UID_i$ does not enable recovery of the master key $MK$, thereby preserving its confidentiality. (2) **User isolation:** The pseudorandomness of HMAC ensures that different user keys are mutually independent, preventing an adversary from deriving one user's key from another user's key. (3) **Key-derivation restriction:** An adversary without knowledge of $MK$ cannot derive a valid user key for an arbitrary user identifier. The master key $MK$ is retained by the model owner and is not included in the distributed materials.

The model owner encodes the byte contents of the obfuscated model and the custom inference library according to a predefined file order and length-prefix convention, and computes their SHA-256 digest $H_M$. The model identifier $MID$, version number $v$, user identifier $UID_i$, and $H_M$ are then used to construct the authenticated associated data:

$$A_i = Encode\left(\text{“KeyObf”}, MID, v, UID_i, H_M\right). \tag{7}$$

Let all $D$ key factors associated with the same model form a key-factor vector $r_{key} = \left(r_{key,1}, \ldots, r_{key,D}\right)$. The model owner then generates a 96-bit random nonce $N_i$ for user $i$, which must not be reused under the same user key. Using the 256-bit user key $UK_i$, the model owner applies AES-256-GCM to perform authenticated encryption of the fixed-length encoding of the key-factor vector:

$$\left(C_i, T_i\right) = AES-256-GCM.Enc\left(UK_i, N_i, Encode\left(r_{key}\right), A_i\right). \tag{8}$$

Here, $C_i$ denotes the ciphertext, and $T_i$ is a 128-bit authentication tag. The function $Encode\left(r_{key}\right)$ encodes the key factors in a fixed order determined by the coupling positions and output-feature indices, together with the vector length and numerical format identifier. The independent authorization credential for user $i$ is defined as

$$L_i = \left(MID, v, UID_i, H_M, N_i, C_i, T_i\right). \tag{9}$$

The authorization credential $L_i$ is generated and issued as a separate file outside the model. The obfuscated model and custom inference library are identical for all users. The model owner distributes $UK_i$ to the corresponding user through a secure channel and provides the associated credential $L_i$. Since $L_i$ contains the encrypted key factors and the metadata required for authentication, it can be stored and distributed as an ordinary file alongside the application, whereas the user key $UK_i$ should be stored in the device's secure key storage facility. When a new user is added, the model owner only needs to derive a user key from the new user identifier, re-encrypt the same key-factor vector, and issue a new authorization credential.

## 3.4 Authorized Inference

Authorized inference is performed by the user on a local device. The inference process can be expressed as

$$Y = F_{M'}\left(X, UK_i, UID_i, L_i\right). \tag{10}$$

Here, $M'$ denotes the obfuscated model, $X$ is the input data for inference, $UK_i$ is the user key, $UID_i$ is the user identifier, and $L_i$ is the user's independent authorization credential. The inference library operates as follows. First, the credential verification module checks the user identifier, model identifier, and version number contained in the credential, computes the SHA-256 digest of the current model package, and constructs the associated data $A_i$ in the same manner as during credential issuance. It then performs AES-256-GCM authenticated decryption using $UK_i$ and the values $N_i$, $C_i$, and $T_i$ contained in the credential:

$$AES-256-GCM.Dec\left(UK_i, N_i, C_i, T_i, A_i\right) = \begin{cases} r_{key}, & \text{if authentication succeeds}, \\ \perp, & \text{if authentication fails}. \end{cases}$$

Upon successful authentication, the inference library checks the length, ordering, factor configuration, and numerical format of the key-factor vector. Inference proceeds only after these properties are confirmed to match the coupling configuration of the current model. Each coupling layer multiplies its output by the corresponding key factor to complete the compensation operation defined in Eq. (3). The model digest is verified by the inference library on the user's device to prevent a credential from being mistakenly used with a different model or version; this verification does not provide a tamper-resistance guarantee.

When the user key is correct and the credential matches both the model and the user, authenticated decryption recovers the key-factor vector $r_{key}$. The effective recovery coefficient at each coupling layer is then $r_{emb,j} \cdot r_{key,j} = 1/\alpha_j$, which exactly compensates for the parameter transformation applied to the selected operator, making the obfuscated model functionally equivalent to the original model. If the user key is incorrect, the credential is missing, or the ciphertext or its associated binding information has been tampered with, authenticated decryption returns the failure state $\perp$. In this case, the inference library does not release the key factors or execute subsequent model inference.

## 3.5 Correctness Analysis

This section analyzes the correctness of the obfuscation scheme introduced in Section 3.2.1 and proves that the parameter obfuscation transformations do not affect the output of authorized inference.

**Proof of Lemma 3.1.** For a purely linear subnetwork, we analyze the transformed outputs layer by layer.

For the first layer, we have

$$X_1' = \left(\alpha W_1\right)^T X_0 + \alpha b_1 = \alpha X_1,$$

which scales the output by a factor of $\alpha$.

For the intermediate layers ($i \in [2, n-1]$; no intermediate layers exist when $n=2$), the weights remain unchanged, while the biases are multiplied by $\alpha$. Thus,

$$X_i' = W_i^T \left(\alpha X_{i-1}\right) + \alpha b_i = \alpha X_i,$$

preserving the scaling relationship.

For the final layer, the weights are multiplied by $1/\alpha$, while the bias remains unchanged. Therefore,

$$X_n' = \left(\frac{1}{\alpha} W_n\right)^T \left(\alpha X_{n-1}\right) + b_n = X_n,$$

which restores the original output. Hence, $f'\left(X_0\right) = f\left(X_0\right)$. QED.

**Proof of Theorem 3.2.** For a subnetwork satisfying the conditions of the theorem, we first establish the propagation of the scaling factor.

The transformed output of the first layer is

$$X_1' = \left(\alpha W_1\right)^T X_0 + \alpha b_1 = \alpha X_1.$$

Assume that the $(i-1)$-th layer satisfies $X_{i-1}' = \alpha X_{i-1}$. Since the weights of the $i$-th layer ($2 \le i \le n$) remain unchanged and its bias is scaled to $\alpha b_i$, we obtain

$$X_i' = W_i^T X_{i-1}' + \alpha b_i = \alpha X_i.$$

Therefore, mathematical induction gives

$$f'(X_0) = \alpha f(X_0).$$

Next, let $z$ be an arbitrary component of $f(X_0)$. When $z \le 0$, both sides produce an output of zero. When $0 < z < \beta$, we have $0 < \alpha z < \beta$, and inverse scaling restores the value to $z$. When $z \ge \beta$, regardless of whether $\alpha z$ reaches the clipping threshold, inverse scaling followed by the outer clipped ReLU produces an output of $\beta$.

Thus, for any $z$,

$$ReLU_\beta\left(\frac{1}{\alpha} ReLU_\beta(\alpha z)\right) = ReLU_\beta(z).$$

Applying this property element-wise to the subnetwork output and combining it with $f'(X_0) = \alpha f(X_0)$ yields Eq. (2). The theorem is therefore proved. QED.

# 4 Experimental Evaluation

## 4.1 Experimental Setup

**Hardware and Software Environment.** Experiments were conducted on a workstation equipped with an Intel Xeon w7-3445 processor, 64 GB of RAM, and an NVIDIA RTX 6000 Ada Generation GPU. The software environment consisted of Ubuntu 24.04, Python 3.7, and TensorFlow 2.x for model construction and inference evaluation. Models were deployed and protected in TFLite format. The key derivation module used Python’s built-in hmac and hashlib libraries to implement deterministic key generation based on HMAC-SHA256.

**Models and Dataset.** Six neural network models were selected for evaluation: VGG11, MobileNetV2, and four ResNet variants (ResNet20, ResNet32, ResNet44, and ResNet56). These models cover a range of architectures, from lightweight to medium-sized networks, allowing us to evaluate the applicability of the proposed scheme. All models were evaluated on the CIFAR-100 dataset, which contains 60,000 images across 100 classes.

**Parameter Configuration.** The scaling factor $\alpha$ for each selected layer was independently sampled from the interval $(0.1, 0.4)$. Key factors were derived using HMAC-SHA256 and mapped to the interval [10,50) through MapToFloat(). They were generated and stored in float64 precision, while inference computations used float32 precision. Each model was configured with 10 coupling positions, with independent key factors assigned to individual output features, corresponding to output channels in convolutional layers. Table 4 further compares the effectiveness of key-factor recovery under different channel-grouping granularities.

## 4.2 Results and Analysis

**Accuracy and Overhead.** Table 1 presents the experimental results of KeyObf on six neural network models. When the correct user key, user identifier, and authorization credential are provided, the Top-1 accuracy of the obfuscated models on CIFAR-100 remains nearly identical to that of the original models, with an average change of approximately −0.01 percentage points. When an incorrect user key is provided or the authorization credential is missing, the inference library rejects authentication and does not execute subsequent model inference. In the inference time evaluation using 10,000 test samples, the average inference time across the six models increases from 208.65 ms to 391.31 ms, with relatively larger increases observed for the ResNet variants. These results indicate that KeyObf preserves the accuracy of authorized inference while introducing additional computational overhead, with inference time remaining on the millisecond scale.

**Table 1. Accuracy and inference time of KeyObf on each model**

| Model | Original Top-1 accuracy (%) | Obfuscated Top-1 accuracy (%) (Δ) | Original inference time (ms) | Obfuscated inference time (ms) (Δ) |
|---|---|---|---|---|
| VGG11 | 70.80 | 70.79 (−0.01) | 124.85 | 195.85 (+71.00) |
| MobileNetV2 | 74.29 | 74.29 (0.00) | 309.61 | 375.25 (+65.64) |
| ResNet20 | 68.83 | 68.77 (−0.06) | 118.66 | 243.33 (+124.67) |
| ResNet32 | 70.14 | 70.15 (+0.01) | 174.32 | 380.98 (+206.66) |
| ResNet44 | 71.66 | 71.62 (−0.04) | 230.80 | 519.68 (+288.88) |
| ResNet56 | 72.61 | 72.63 (+0.02) | 293.66 | 632.77 (+339.11) |
| **Average** | **71.39** | **71.38 (−0.01)** | **208.65** | **391.31 (+182.66)** |

**Comparison with Existing Methods.** Table 2 compares KeyObf with ModelObfuscator and DynaMO, and further evaluates the independent contribution of key binding through an ablation experiment. In terms of usage authorization, existing methods allow normal inference without credentials, whereas both KeyObf and its variant without key binding reject inference when the credential is missing. When authentication is bypassed under controlled conditions, the variant without key binding can still perform normal inference, whereas the complete KeyObf scheme cannot perform correct compensation because the key factors are unavailable. For dynamic analysis, we use the dynamic instrumentation tool DLModelExplorer to evaluate the extractability of model weights. KeyObf achieves a weight extraction rate of 48.9%, close to the 51.6% achieved by DynaMO, and this rate remains unchanged with or without key binding. In terms of inference overhead, key binding increases the average inference time from 390.01 ms to 391.31 ms, an increase of approximately 0.33%. These results indicate that key binding introduces an additional barrier to restoring model functionality while largely preserving the original obfuscation effectiveness and inference performance.

**Table 2. Comparison between KeyObf and existing model obfuscation methods (CIFAR-100)**

| Method | Coupling transformation | Without credential | After authentication bypass | Weight extraction rate (%) | Average inference time (ms) |
|---|---|---|---|---|---|
| Original model | No | Normal inference | — | 100.00 | 208.65 |
| ModelObfuscator | No | Normal inference | — | 97.76 | 335.00 |
| DynaMO | Yes | Normal inference | — | 51.60 | 388.00 |
| KeyObf (without key binding) | Yes | Inference rejected | Normal inference | 48.90 | 390.01 |
| KeyObf | Yes | Inference rejected | Invalid output | 48.90 | 391.31 |

*Note: "—" indicates that the method has no authentication mechanism. Inference times are averaged across the six models.*

**Evaluation of Resistance to Reverse Engineering.** To evaluate the effectiveness of KeyObf against model reverse engineering, we consider three common attack methods: (1) *Model format conversion*, which attempts to recover the model by exporting the obfuscated model into a standard, parsable format using conversion tools such as tf2onnx, tflite2onnx, and tflite2tf; (2) *Buffer reverse parsing*, which attempts to reconstruct the original model architecture by parsing the FlatBuffer serialization structure; and (3) *Feature-behavior analysis*, which attempts to infer the model structure and replicate its functionality from input-output behavior. As shown in Table 3, under the evaluated conditions, general-purpose model conversion and buffer parsing tools fail to directly recover the original model. In model format conversion attacks, operator renaming prevents conversion tools from recognizing standard operator types, causing the conversion process to terminate or produce invalid files. In buffer reverse parsing attacks, the obfuscated serialization structure disrupts the original field mappings, preventing attackers from extracting meaningful architectural information. In feature-behavior analysis attacks, nonlinear distortion of the parameter space prevents attackers from inferring the internal structure from input-output behavior.

**Table 3. Defense effectiveness against reverse engineering attacks**

| Category | Tool | Original model | KeyObf | KeyObf defense mechanism |
|---|---|---|---|---|
| Model format conversion | tf2onnx | × | √ | Operators not recognized |
| Model format conversion | tflite2onnx | × | √ | Operators not recognized |
| Model format conversion | tflite2tf | × | √ | Operators not recognized |
| Buffer reverse parsing | FlatBuffer parsing | × | √ | Field mappings disrupted |
| Feature-behavior analysis | Behavioral cloning | × | √ | Invalid inference outputs |

*Note: √ indicates successful defense, whereas × indicates failed defense.*

**Sensitivity to Key-Factor Deviations.** To investigate how deviations in the key factors $r_{key}$ affect model inference, we use ResNet56 to compare two configurations in which the coupling layers are placed either within residual branches or on the main residual path. Taking the correct key-factor vector as the reference, we apply the same relative perturbation to all factors, $r'_{key} = r_{key}(1+\delta)$, where $\delta$ denotes the relative deviation. The experiment uses 1,000 test images, with an original model accuracy of 69.5%. We define 90% of this accuracy (62.55%) as the threshold for effective inference. As shown in Fig. 1, when the coupling layers are placed within residual branches, the range of deviations over which model accuracy reaches the threshold is approximately −20.2% to +20.4%. When the coupling layers are placed on the main residual path, this range narrows to −4.8% to +3.2%, indicating that coupling-layer placement has a substantial effect on the model's sensitivity to key-factor deviations. The subsequent factor search experiments adopt the main-path configuration to evaluate factor recovery under a narrower deviation tolerance.

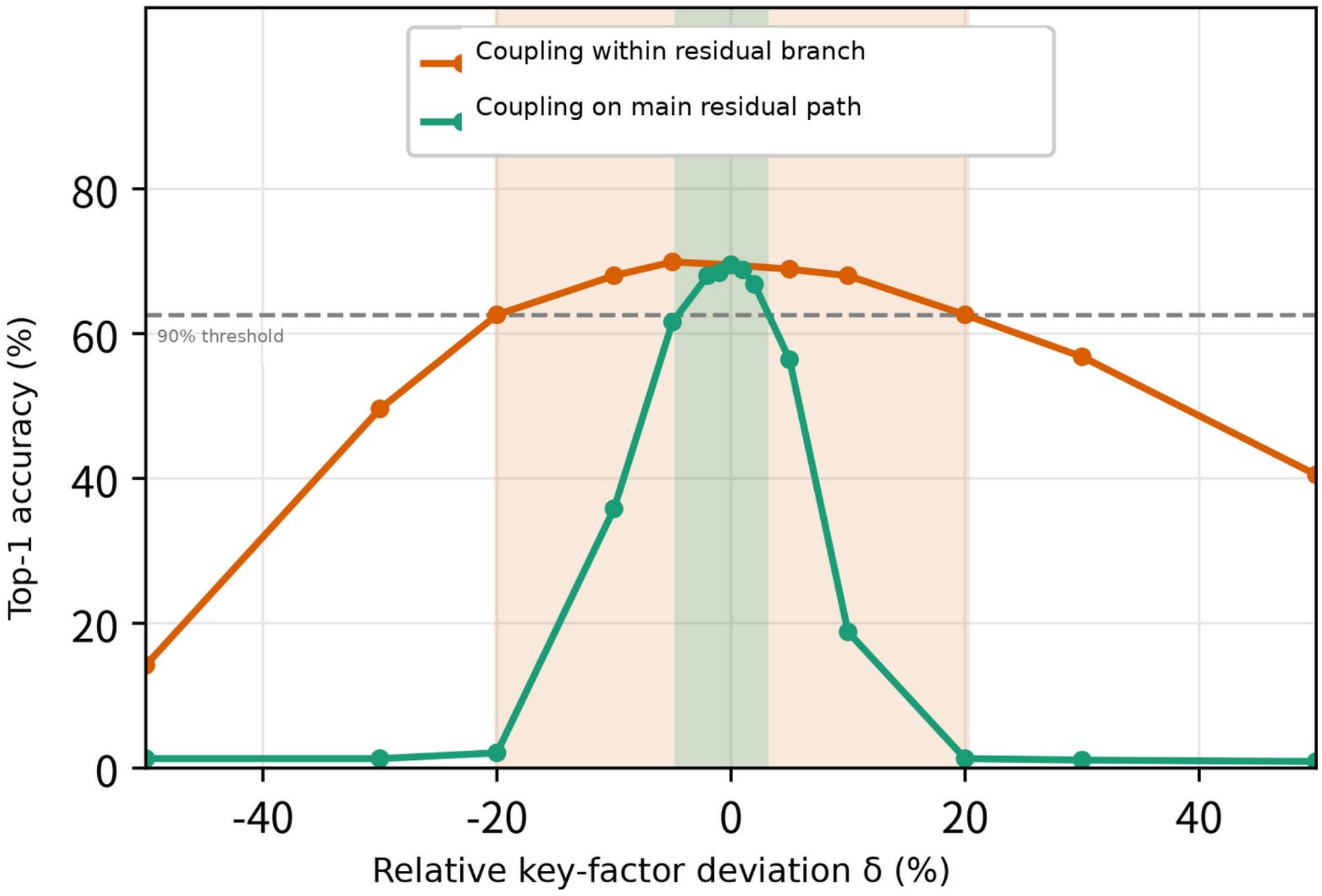

*Figure 1. Accuracy versus relative deviation of key factors (ResNet56).*

**Recoverability Analysis.** KeyObf controls model inference through credential authentication, but the recoverability of key factors requires further evaluation. We conduct a loss-feedback-based black-box factor search using ResNet56. The experiment assumes that the attacker knows the factor range [10,50) and the number of factors $D$ to be recovered, but does not know the correct key factors or the master key. The test model allows candidate factors to be substituted, and cross-entropy loss is computed using 100 labeled samples (one per class, with no overlap with the images used for final evaluation). Each loss evaluation for a candidate factor vector is counted as one query. For $D=1$, random search and grid search are employed, while the remaining configurations use the Covariance Matrix Adaptation Evolution Strategy (CMA-ES). Each configuration is allowed a maximum of 2,000 queries and is evaluated over three independent trials. An attack is considered successful if the recovered model achieves at least 90% of the original model's accuracy. As shown in Table 4, low-dimensional factor configurations can recover relatively high inference accuracy within a limited number of queries. In particular, the per-position design ($D=10$) achieves a success rate of 3/3. As the number of factors increases, recovery performance generally declines. When $D \geq 88$, none of the trials reaches the recovery threshold. These results indicate that fine-grained factor configurations increase the difficulty of recovering model functionality under the evaluated black-box attack conditions.

**Table 4. Recovery results of key factors under different designs (ResNet56, CIFAR-100)**

| Factor design | Number of factors $D$ | Attack method | Queries | Successful trials / total trials | Mean recovered accuracy (%) (range) |
|---|---|---|---|---|---|
| Globally shared | 1 | Random search | 50 | 0/3 | 5.5 (0.8–8.7) |
| Globally shared | 1 | Grid search | 53 | 3/3 | 71.1 (70.7–71.6) |
| Per-position | 10 | CMA-ES | 2,000 | 3/3 | 68.2 (65.6–69.7) |
| Shared per 16 channels | 22 | CMA-ES | 2,000 | 1/3 | 63.9 (60.4–66.2) |
| Shared per 8 channels | 44 | CMA-ES | 2,000 | 0/3 | 52.6 (48.2–55.1) |
| Shared per 4 channels | 88 | CMA-ES | 2,000 | 0/3 | 7.2 (1.9–10.3) |
| Shared per 2 channels | 176 | CMA-ES | 2,000 | 0/3 | 1.6 (1.2–2.1) |
| Per-channel | 352 | CMA-ES | 2,000 | 0/3 | 1.9 (1.2–2.9) |

## 5 Conclusion

This paper proposes KeyObf, a lightweight model obfuscation and controllable authorization scheme, to address the challenges of model asset protection and usage authorization in third-party deployment scenarios of DLaaS. By binding the recovery coefficients of coupling transformations to user keys, the proposed scheme integrates model protection with usage authorization. Experimental results demonstrate that KeyObf preserves the accuracy of authorized inference, prevents unauthorized inference when credential authentication fails, and provides a certain degree of model protection against the evaluated reverse-engineering attacks. Furthermore, the factor recovery experiments show that the granularity of key-factor configuration has a significant impact on the difficulty of recovering model functionality, with finer-grained factor designs exhibiting greater resistance to search-based attacks under the given query budget. Future work will further investigate mechanisms for protecting and dynamically managing key factors in local execution environments, including key revocation, online updates, and fine-grained authorization policies based on time, usage count, or functional modules, to support more complex model distribution requirements.